# Comment on "A de Haas-van Alphen study of the Fermi surfaces of superconducting LiFeP and LiFeAs"

S.V. Borisenko,[1] V. B. Zabolotnyy,[1] D.V. Evtushinsky,[1] T. K. Kim,[1,2] I.V. Morozov,[1,3] A. A. Kordyuk,[1] B. Büchner[1]

[1] *Leibniz-Institute for Solid State Research, IFW-Dresden, D-01171 Dresden, Germany*
[2] *Diamond Light Source Ltd, Oxfordshire OX11 0DE, UK*
[3] *Moscow State University, Moscow 119991, Russia*

In a recent preprint[1] Putzke et al. argued that their dHvA data on LiFeAs are in good agreement with DFT calculations and contradict our ARPES results[2]. Here we show that the situation is just the opposite.

In left panel of Fig.1 we reproduce the comparison suggested by Putzke et al. Only three out of ten base frequencies predicted by DFT are observed and none of them match. Having no information about the larger Fermi surface sheets the authors focus their attention on the smaller ones for which, as found later in the paper, the experimental Fermi velocities are up to five times lower than theoretical.

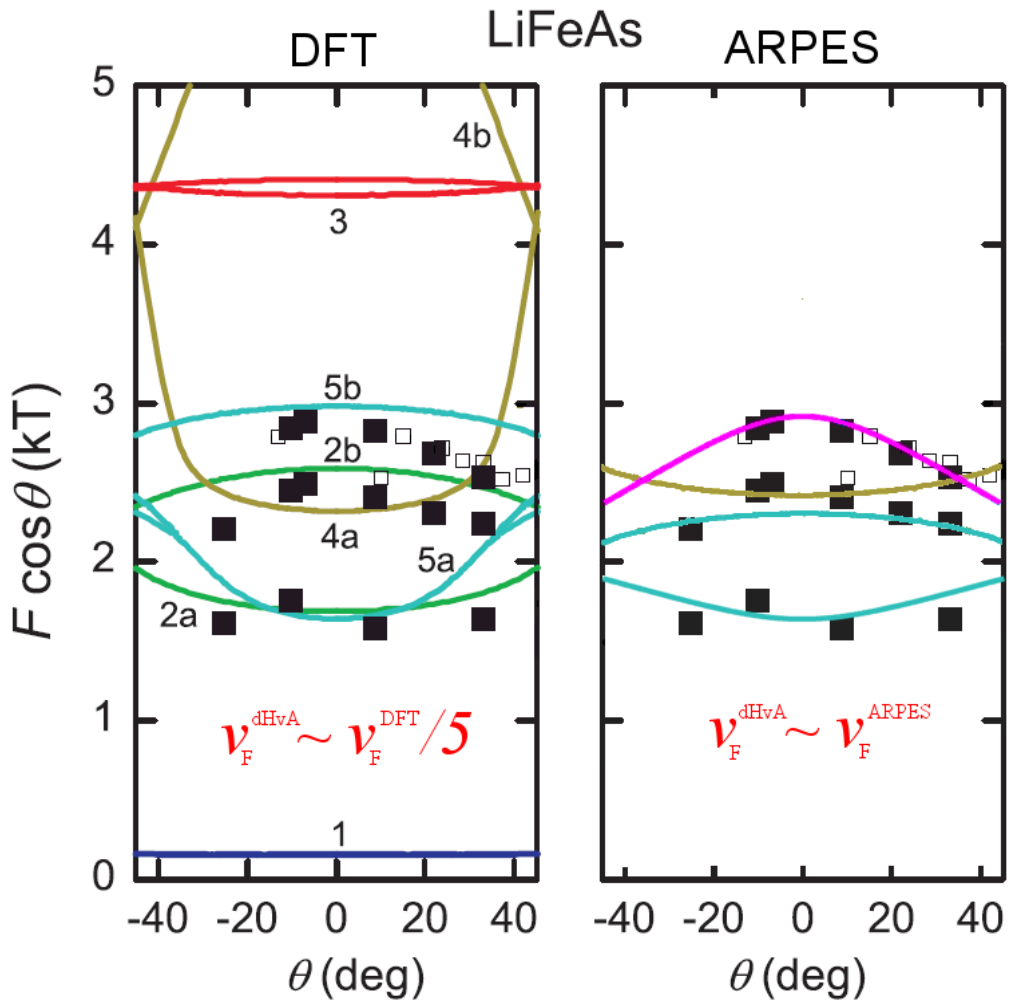

Fig.1 Comparison of the dHvA results with DFT (from Ref.1) and ARPES. Frequencies at θ=0 are determined by external areas of Fermi surfaces measured in a wide range of excitation energies. ARPES curves are taken from the calculations (see Fig.2 right panel), but are rescaled in accordance with the experimental degree of warping. Apart from the shown data, there is a maximum of outer electron pocket at 3660T. This orbit is not seen in dHvA presumably either because it is too long or because it is not smooth enough. The largest hole pocket is very 2D and has an area of 5500T. The middle hole pocket is very small (< 350T) and has a $k_z$ dispersion of the order of 10 meV. One might also expect to see a minimum of the ellipse (~2000T) and even other hybrid orbits originated from the crossed ellipses (see inset to Fig.3).

Because of intrinsic inability of the method to determine either the location in the Brillouin zone or the topology of the given Fermi surface sheet they apply a special procedure to obtain nesting conditions. Virtually unlimited possibilities to tailor a required electronic structure are offered by shifting the calculated bands independently by an arbitrary energy. In the case of LiFeAs, however, the five Fermi surfaces cannot be described by five independent dispersing features. For instance, shifting the one which supports the hole pocket would imply the same shift of a corresponding electron-like feature[2]. This is apparently not an obstacle for Putzke et al. as they do shift DFT hole- and electron-like features independently, such that both shrink, quoting the numbers (up to 73 meV for LiFeP! ) with an elusive physical meaning (in view of 400% mismatch of Fermi velocities). Even such an unjustified shaping of DFT results is obviously not successful as the right panel in Fig. 2c of Ref.1 suggests: base frequencies at $\theta = 0$ are still not reproduced and $\theta$ - dependences (curvatures or angular dispersions) of $\varepsilon$ and $\gamma$ orbits remain considerably stronger in the experiment. Furthermore, the question as to why the minimum of the outer electron pocket (4a) is not observed in LiFeAs, whereas it is seen for the full range of angles in LiFeP, remains open. This clearly poor agreement is called in Ref.1 an "exact match".

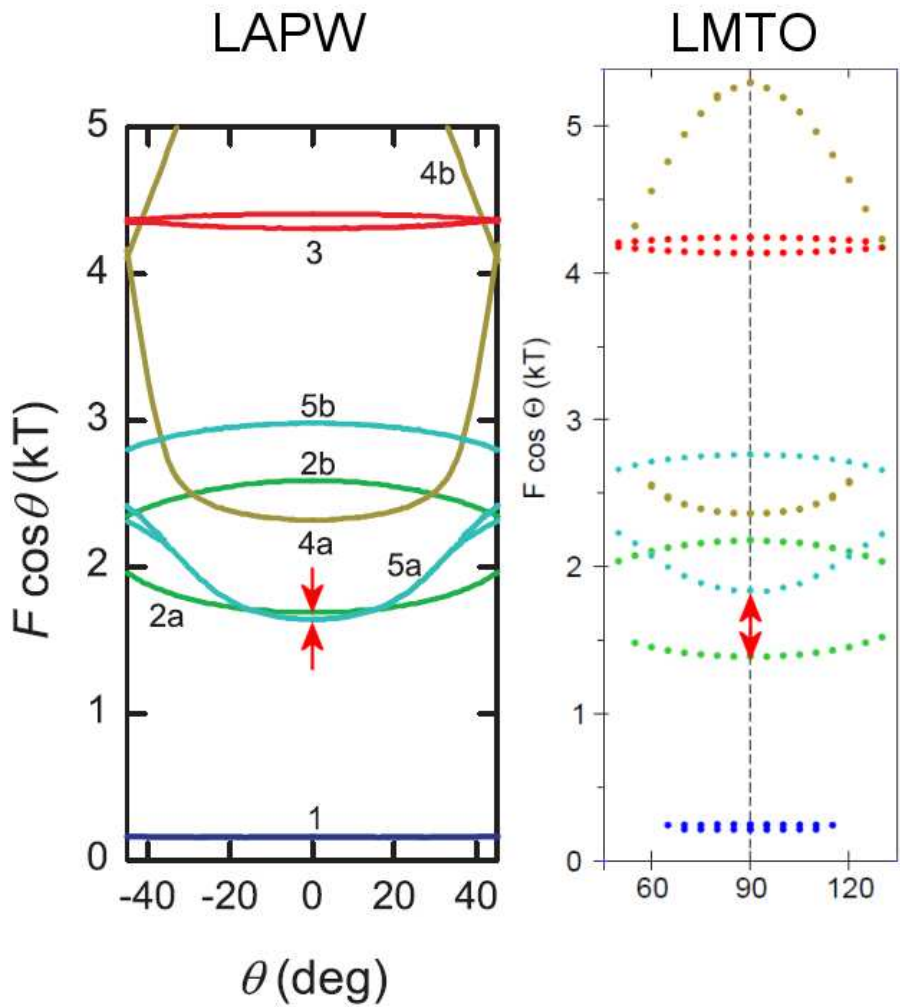

Fig.2 Two versions of the theoretical dHvA angular dispersions from band structure calculations. Left panel is adapted from Ref.1. Right panel – LMTO result. Red arrows show that degeneracy of the 2a and 5a orbits is not universal and depends on the computational scheme.

The interpretation offered by Putzke et al. is fully based on the accidental near-degeneracy of the 5a and 2a frequencies (green and blue curves in Fig.2c of Ref.1). Calculations of Yaresko et al., shown in Fig.2 together with the results of commercial WIEN2K code used in the preprint, demonstrate that this degeneracy is indeed not universal. In LMTO calculations case the size of the middle hole pocket is noticeably smaller as is the warping of inner electron pocket (blue points), which is the correct trend seen in ARPES experiments[2]. Thus, another frequency is apparently missing in dHvA data.

A more transparent alternative is to compare the results of two experiments directly. In the right panel of Fig.1 we show both the results of dHvA and some of the Fermi surface areas seen by ARPES in this range of frequencies. The base frequencies have been determined from the Fermi surface maps taken at different excitation energies[2,3]. The curvatures are simulated considering the actual (experimental) degree of warping and calculations. For example, the magenta curve originating from the elliptical orbit has ~2.4 times weaker angular dispersion than the brown curve at ~5 kT in right panel of Fig.2.

The nearly perfect agreement with our ARPES data is achieved by considering the elliptical electron-like orbit excluded from the analysis in Ref.1. According to the calculations, the spin-orbit coupling lifts the degeneracy between the elliptical electron-like pockets along the high-symmetry directions of the folded BZ resulting in two separated inner and outer electron pockets. Our experimental data presented in Fig. 3 convincingly demonstrate that this effect is clearly overestimated by the calculations, at least by the order of magnitude. The radial cuts in the vicinity of 90° direction show practically a single dispersing feature. This means that together with the inner and outer electron pockets one should consider ellipses as possible orbits (inset to Fig.3), and they, being doubled and smooth, are expected to give the strongest dHvA signal, as is the case (Fig.1 of Ref.1). Even if a tiny gap formally exists, it is difficult to estimate the breakdown field from either $\delta k$ (criterion used in Ref.4) or $\delta e$ (criterion used in Ref.5) since we deal here not with an ideal band structure with the delta-function-like spectral function, but with a spectral function which has an intrinsic (non-resolution limited) width at ($E_F$, $k_F$) even being measured at T~1K (Fig.3). Considering the 80 Tesla breakdown field for 30 meV splitting estimated by some of the authors of Ref.1 for the ideal band structure of 1111 compound[4] it is clear that in the current situation even a formal estimate would yield much lower values than the typical magnetic fields used in Ref.1.

Our interpretation thus suggests that two experiments agree within the error bars, which implies absence of nesting in LiFeAs, as has been shown by us earlier[2]. This agreement is not surprising since in the case of LiFeAs (no polar surface!) ARPES delivers very detailed and bulk-representative information about the low energy electronic structure. This is expected from theory[6] and has been confirmed by comparison with other experiments[7,8,9]. Such an agreement with another bulk-sensitive technique gives just another evidence for that. Still somewhat lower frequencies (e.g. minimum of inner electron pocket in Fig.1 right panel) could be related with the Li-deficiency (effective hole doping) and thus different handling of the single crystals known to be very reactive. Our samples spent less than a minute on open air prior to be cleaved and measured in ultra high vacuum.

In spite of very limited information about the Fermi surface delivered by dHvA experiments, they still can be useful when studying the electronic structure of 3D materials. Although the finite integration over $k_Z$ is not an obstacle for ARPES to determine the areas of the extremal cross sections, the information about their angular dispersion is uniquely available from dHvA studies. However, as the example above demonstrates, without a direct comparison with ARPES experiment, this information can be easily misinterpreted.

We conclude by noting, that all the quantitative estimates we made from ARPES measurements including the information on the elliptical orbits had been communicated to the authors *before* we became aware of the results of dHvA studies. Nevertheless, Putzke et al. have chosen to interpret their results as being in a good agreement with DFT.

We are grateful A. Yaresko for sharing with us the results of LMTO band structure calculations. We thank A. Coldea and A. Carrington for discussions. The work is supported by DFG grant "Iron pnictides".

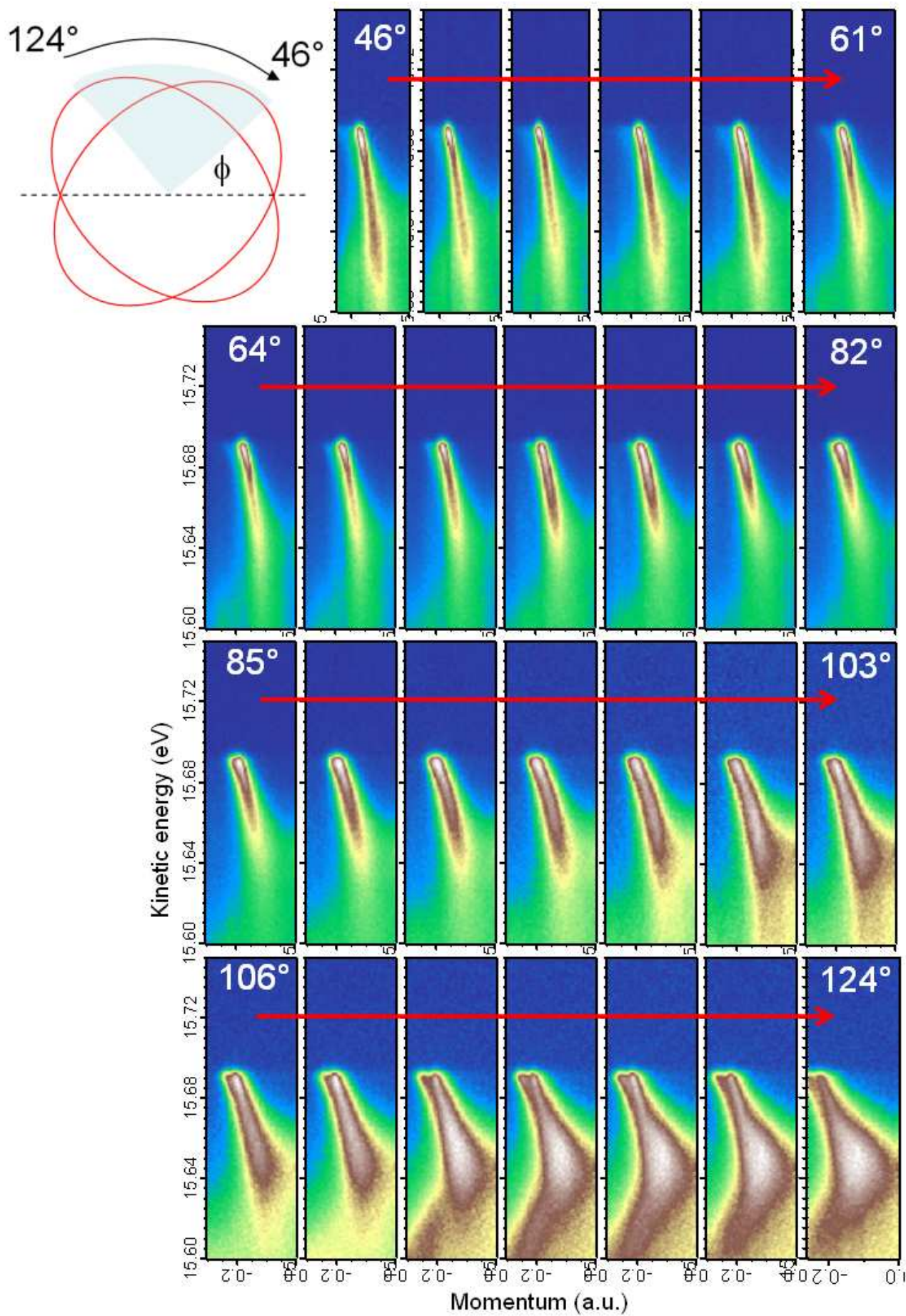

Fig. 3 Radial cuts of the electronic structure from the corner of the folded Brillouin zone as seen in ARPES experiment. Note, that the cuts at e.g. 124° and 55° are not equivalent because of the matrix element effect which changes the ratio of photoemission intensities of two features.